
\magnification = 1200
\font\eightrm=cmr8
\font\eighti=cmmi8
\font\eightsy=cmsy8
\font\eightbf=cmbx8
\font\eighttt=cmtt8
\font\eightit=cmti8
\font\eightsl=cmsl8
\font\sixrm=cmr6
\font\sixi=cmmi6
\font\sixsy=cmsy6
\font\sixbf=cmbx6
\catcode`@11
\newskip\ttglue
\font\grrm=cmbx10 scaled 1200

\def\eightpoint{\def\rm{\fam0\eightrm}
\textfont0=\eightrm \scriptfont0=\sixrm \scriptscriptfont0=\fiverm
\textfont1=\eighti \scriptfont1=\sixi \scriptscriptfont1=\fivei
\textfont2=\eightsy \scriptfont2=\sixsy \scriptscriptfont2=\fivesy
\textfont3=\tenex \scriptfont3=\tenex \scriptscriptfont3=\tenex
\textfont\itfam=\eightit \def\it{\fam\itfam\eightit}
\textfont\slfam=\eightsl \def\sl{\fam\slfam\eightsl}
\textfont\ttfam=\eighttt \def\tt{\fam\ttfam\eighttt}
\textfont\bffam=\eightbf
\scriptfont\bffam=\sixbf
\scriptscriptfont\bffam=\fivebf \def\bf{\fam\bffam\eightbf}
\tt \ttglue=.5em plus.25em minus.15em
\normalbaselineskip=6pt
\setbox\strutbox=\hbox{\vrule height7pt width0pt depth2pt}
\let\sc=\sixrm \let\big=\eightbig \normalbaselines\rm}
\newinsert\footins
\def\newfoot#1{\let\@sf\empty
  \ifhmode\edef\@sf{\spacefactor\the\spacefactor}\fi
  #1\@sf\vfootnote{#1}}
\def\vfootnote#1{\insert\footins\bgroup\eightpoint
  \interlinepenalty\interfootnotelinepenalty
  \splittopskip\ht\strutbox 
  \splitmaxdepth\dp\strutbox \floatingpenalty\@MM
  \leftskip\z@skip \rightskip\z@skip
  \textindent{#1}\footstrut\futurelet\next\fo@t}
\def\fo@t{\ifcat\bgroup\noexpand\next \let\next\f@@t
  \else\let\next\f@t\fi \next}
\def\f@@t{\bgroup\aftergroup\@foot\let\next}
\def\f@t#1{#1\@foot}
\def\@foot{\strut\egroup}
\def\footstrut{\vbox to\splittopskip{}}
\skip\footins=\bigskipamount 
\count\footins=1000 
\dimen\footins=8in 

\def\ref#1{$^{#1}$}
\def\flex{\raise 6pt\hbox{$\leftrightarrow $}\! \! \! \! \! \! }
\def\tr{ \mathop{\rm tr}}

\newbox\bigstrutbox
\setbox\bigstrutbox=\hbox{\vrule height10pt depth5pt width0pt}
\def\bigstrut{\relax\ifmmode\copy\bigstrutbox\else\unhcopy\bigstrutbox\fi}
\def\refer[#1/#2]{ \item{#1} {{#2}} }
\def\rev<#1/#2/#3/#4>{{\it #1\/} {\bf#2}, {#3}({#4})}
\def\boxit#1{\vbox{\hrule\hbox{\vrule\kern3pt
\vbox{\kern3pt#1\kern3pt}\kern3pt\vrule}\hrule}}

\def\2figure#1#2#3#4{\vbox{ \hrule width#1truecm \hbox{\vrule height#2truecm
\hskip #1truecm
\vrule height#2truecm }\hrule width#1truecm \hbox{\vrule\vbox{\hsize #1truecm
\baselineskip=10pt
\noindent\strut#3}\vrule}\hrule width#1truecm
\hbox{\vrule\vbox{\hsize #1truecm
\baselineskip=10pt
\noindent\strut#4}\vrule}\hrule width#1truecm  }}
\def\3figure#1#2#3#4#5{\vbox{ \hrule width#1truecm \hbox{\vrule height#2truecm
\hskip #1truecm
\vrule height#2truecm }\hrule width#1truecm \hbox{\vrule\vbox{\hsize #1truecm
\baselineskip=10pt
\noindent\strut#3}\vrule}\hrule width#1truecm
 \hbox{\vrule\vbox{\hsize #1truecm
\baselineskip=10pt
\noindent\strut#4}\vrule}
\hrule width#1truecm \hbox{\vrule\vbox{\hsize #1truecm
\baselineskip=10pt
\noindent\strut#5}\vrule}\hrule width#1truecm  }}

\def\sqr#1#2{{\vcenter{\hrule height.#2pt
   \hbox{\vrule width.#2pt height#1pt \kern#1pt
    \vrule width.#2pt}
    \hrule height.#2pt}}}


\font\sf=cmss10

\def\smin{\,\raise 0.06em \hbox{${\scriptstyle \in}$}\,}
\def\smsubset{\,\raise 0.06em \hbox{${\scriptstyle \subset}$}\,}

\def\Natural{\hbox{\hskip 1.5pt\hbox to 0pt{\hskip -2pt I\hss}N}}
\def\Integer{\>\hbox{{\sf Z}} \hskip -0.82em \hbox{{\sf Z}}\,}
\def\Rational{\hbox{\hbox to 0pt{\hskip 2.7pt \vrule height 6.5pt
                                  depth -0.2pt width 0.8pt \hss}Q}}
\def\Real{\hbox{\hskip 1.5pt\hbox to 0pt{\hskip -2pt I\hss}R}}
\def\Complex{\hbox{\hbox to 0pt{\hskip 2.7pt \vrule height 6.5pt
                                  depth -0.2pt width 0.8pt \hss}C}}
\def \E {{{\rm e}}}
\def \dum {\partial ^{{}^{\!\!-\!1}}\!\!\!}
\def \0j {j_{{}_0} }
\def \1j {j_{{}_1} }
\def \sq {\overline Q }
\def \J {{\cal J} }
\def \ss#1{{\scriptstyle#1}}
\centerline{\grrm The Algebra of Non-Local Charges in Non-Linear Sigma Models}
\vskip 1.5cm
\centerline { E. Abdalla$^{(1)}$\newfoot {${}^*$}{Partially supported by CNPq.
Address after August/93: CERN-Theory Division, CH-1211, Gen\`eve 23,
Switzerland.},
M.C.B. Abdalla$^{(2)}{}^*$, J.C. Brunelli$^{(2)}$\newfoot {${}^\dagger$}
{Supported by CNPq.} and A. Zadra$^{(1)}$\newfoot {${}^\star$}{Supported by
FAPESP.}}
\vskip 1.5truecm
\centerline {${}^{(1)}$ Instituto de F\'\i sica da  Universidade de
S\~ao Paulo,}
\centerline{Departamento de F\'\i sica Matem\'atica, Caixa Postal 20516,}
\centerline{CEP 01498-970, S\~ao Paulo, SP, Brazil}
\vskip .5truecm
\centerline{${}^{(2)}$ Instituto de F\'\i sica Te\'orica, Universidade Estadual
Paulista,}
\centerline{Rua Pamplona 145, CEP 01405-900, S\~ao Paulo, SP, Brazil}
\vskip 2cm
\centerline{\bf Abstract}
\vskip .5cm

\noindent We obtain the exact Dirac algebra obeyed by the conserved non-local
charges in bosonic non-linear sigma models. Part of the computation is
specialized for a symmetry group $O(N)$. As it turns out the algebra
corresponds to a cubic deformation of the Kac-Moody algebra. The non-linear
terms are computed in closed form. In each Dirac bracket we only find highest
order terms (as explained in the paper), defining a saturated algebra. We
generalize the results for the presence of a Wess-Zumino term. The algebra is
very similar to the previous one, containing now a calculable correction of
order one unit lower.

\vfill \eject
\noindent {\bf 1. Introduction}
\vskip 1.truecm

\noindent In general, quantum field theoretic models where non-perturbative
computations are\break
known, contain an infinite number of conservation laws [1,2]. In fact, the
solvability of several exact $S$-matrices in two dimensional
models can be traced back to the Yang-Baxter relations [3,4], which in turn are
a direct consequence of the conservation of higher powers of the momentum.
Alternatively, there is an infinite number of non-local conservation laws
in most of these models as well [2,5]. Both sets of conserved quantities
can  be related to the  existence of a
Lax pair in the theory: demanding compatibility of the Lax pair, one arrives at
conserved charges as functions of the so called spectral parameter implying,
after Taylor expansion, an infinite number of conservation laws.

Another  set of models containing an infinite number of conserved
quantities are  the two dimensional conformally invariant theories [6,7].
The Virasoro
generators are a generalization of the energy momentum conserved charges.
Defining a realization of the symmetry in terms of the null vectors implies
a number of differential equations to be obeyed by the correlation functions
which can be integrated.  In other words, a further knowledge of the underlying
algebra
obeyed by the conserved quantities, namely the Virasoro algebra, together with
the differential representation of the conserved charges, permitted one to go
one step further, i.e. the complete computation of the correlators.

Our aim here is to obtain the algebra of conserved quantities for
integrable theories. The algebra of local conservation laws is abelian and
therefore  too simple.
Massive perturbations of the conformal generators are also a possibility, since
they also form a non-commuting algebra, and it would be worthwhile to
understand the algebra, as well as the role played by the conservation laws
surviving the mass perturbation [8]. For free fermions ($k=1 $ WZW models)
the results conform to our expectation [9].

Non-local conserved charges, on the other hand, are very powerful objects.
The first non-trivial one alone fixes almost completely the on-shell
dynamics [5,10].

Infinite algebras connected with non-trivial conserved quantities could thus be
the key ingredient for the complete solvability of integrable models, and the
knowledge of their correlation functions. It is thus no wonder that the problem
evaded solution in spite of several attempts. Indeed, it has been claimed long
ago [11] that non-local charges might build up a Kac-Moody algebra, but the
appearance of cubic terms found by several authors showed that the algebraic
problem was much more involved [12-14]. For non-linear sigma models with  a
simple gauge group the quantum non-local charges present no anomaly [15],
and the monodromy
matrix can be computed. Therefore the non-local charge algebra should be
manageable; however, as it turns out, the complete algebra was not known, and
one had hints that a possible break of the Jacobi identity might occur [12].

We show that there is a natural recombination of the standard
non-local charges, whose algebra has an approachable structure, being composed
of a linear part of the Kac-Moody form, and a calculable cubic term. Later we
add a Wess-Zumino (WZ) term to the action, and show that both linear and cubic
pieces of the algebra acquire a further contribution.

In order to find these results we adopt the following strategy. First we
compute explicitly the first few conserved  charges
generated by the procedure of Br\'ezin et al. [16]: the
Dirac brackets of those charges are rather obscure, as
we compute (there are also examples in the literature [12-14]). Therefore we
subsequently define an {\it improved} set of
charges in order to simplify the algebra. By inspection, we propose an Ansatz
for the general algebra of the improved charges. Then we argue, based on the
Jacobi identity proved in the subsequent section, that once we
have verified the algebra up to some order, there must be a set of
charges whose algebra agrees with the Ansatz.

In order to verify the Jacobi identity, we introduce a set of (non-conserved)
charges whose algebra is isomorphic to the Ansatz. In that case it is useful
to start from the analysis of a kind of {\it chain  algebra},
in the sense that we
commute elements defined by  chains  of  local currents tied by a
non-local function in space. In terms of these objects we define a linear
algebra, albeit with a much larger set of terms. Finally, by a sort of trace
projection,  we recover the original  algebra in terms of the {\it saturated}
charges, proving
the Jacobi identity in an indirect way.

This paper is divided as follows: in Sect. 2 we review the algebra obeyed by
Noether local currents based on Refs. [17,18]. In Sect. 3
we consider the canonical
 construction of higher non-local conservation laws and calculate some
non-local charges explicitly. Then we derive the brackets for some
appropriate
combinations of charges and write the Ansatz of the complete algebra. In
Sect. 4 we define the  algebra of saturated charges, which turns out to be
 isomorphic to the algebra of conserved charges. We derive the chain  algebra
structure, the corresponding Jacobi identity, and relate the results to the
case of non-local charges. Based on the chain algebra and the consequent Jacobi
identity, we complete the proof about the highest order structure of the
algebra of improved charges, outlined in Sec. 3.  In Sect. 5 we
introduce the WZ interaction to derive
the corresponding algebra. We leave Sect. 6 for conclusions.

\vskip 1.5truecm
\penalty-300
\noindent {\bf 2. Current Algebra of Non-Linear Sigma Models}
\vskip 1.truecm
\nobreak

\noindent The current algebra of classical non-linear sigma models on
arbitrary Riemannian manifolds $(M)$ is known [17]. Indeed, consider a
non-linear
sigma model on $M$, with metric $g_{ij}(\varphi)$, and the maps $\varphi^i(x)$
from two dimensional Minkowski space $\Sigma$ to $M$.
The sigma model action is given by
$$
S={1\over 2\lambda^2}\int_\Sigma d ^2 x \, \eta^{\mu\nu} g_{ij}(\varphi)
\partial _\mu \varphi^i \partial_\nu \varphi^j\quad .\eqno(1)
$$
The phase space consists of pairs $(\varphi^i(x),\pi_i(x))$, where $\pi$ is a
section of the pull-back $\varphi^*(T^*M)$ of the cotangent bundle of $M$ to
the Minkowski space via $\varphi$, and the canonical equal-time Poisson
brackets read
$$
\eqalign{
\{ \varphi^i(x) , \varphi^j(y)\} &  = \{ \pi_i (x) ,\pi_j(y)\}=0 \cr
\{ \varphi^i(x), \pi_j(y)\} &= \delta^i_j\delta(x-y)\quad .\cr}\eqno(2)
$$
{}From the action (1) we find the canonically conjugated momenta, given by the
expression
$$
\pi_i = {1\over \lambda^2} g_{ij}\dot \varphi^j\quad .\eqno(3)
$$
We suppose that there is a connected Lie group $G$ acting on $M$ by isometries,
such that a generator of the Lie algebra ${\bf g}$ of $G$ is represented by a
fundamental vector field
$$
X_M(m) = {d\over d t} \E^{tX }\cdot m \Big\vert_{t=0} \eqno(4)
$$
on $M$; the Noether current may be defined as
$$
\left( j_\mu ,X\right) =- \left( {1\over \lambda^2}g_{ij} (\varphi)
\partial _\mu \varphi^i X_M^j(\varphi)\right) \quad .\eqno(5)
$$
We define also the symmetric scalar field $j$ as
$$\left( j,X\otimes Y\right)= {1\over \lambda^2} g_{ij}(\varphi)X^i_M Y^j_M
\quad .\eqno(6)
$$
In terms of a basis $t^a$ of ${\bf g}$, such that $[t^a, t^b]=f^{abc}t^c $,
we have
$$
\eqalign{
j_\mu & = j_{\mu }^a t^a\cr
j &= j^{ab}t^a \otimes t^b\quad ,\cr}\eqno(7)
$$
and we find the current algebra
$$
\eqalign{
\{ \0j ^a(x), \0j ^b(y)\}&= -f^{abc}\0j ^c(x)\delta(x-y) \cr
\{ \0j ^a(x), \1j ^b(y)\}&= -f^{abc}\1j ^c(x)\delta(x-y) + j^{ab}(y)
\delta'(x-y)\cr
\{ \1j ^a(x), \1j ^b(y)\}&= 0 \cr
{\hfill }&{\hfill }\cr
\{ \0j ^a(x), j^{bc}(y)\}&= -(f^{abd}j^{cd}(x)+ f^{acd}j^{bd}(x))\delta(x-y)
\cr
\{ \1j ^a(x), j^{bc}(y)\}&= 0 \cr
\{ j^{ab}(x), j^{cd}(y)\}&= 0\quad. \cr}
 \eqno(8)
$$
In order to give explicit examples, although without loss of generality, we
specialize to the $O(N)$ case, with Lagrangian
$$
{\cal L} = {1\over 2}\partial _\mu \varphi_i\partial^\mu \varphi_i \quad ,\quad
\sum _{i=1}^N \varphi^2_i =1\quad ,\eqno(9)
$$
and Hamiltonian density
$$
{\cal H} = {1\over 2}(\pi_i^2 + {\varphi'_i}^2)\quad ,\eqno(10)
$$
where $\pi_i = \dot\varphi_i$. We have to impose the constraints
$$
\varphi^2_i -1 =0 \qquad {\rm and } \qquad \varphi_i\pi_i =0 \quad .\eqno(11)
$$
Dirac brackets can be easily calculated and read
$$\eqalign{
\{\varphi_i(x) , \varphi_j(y) \} & =  0\quad ,\cr
\{\varphi_i(x) , \pi_j(y) \} & =  (\delta_{ij} -\varphi_i\varphi_j) (x)
\delta(x-y)\quad ,\cr
\{ \pi_i(x), \pi_j(y)\} & = -(\varphi_i\pi_j - \varphi_j \pi_i )(x)\delta (x-y)
\quad .\cr }\eqno(12)
$$
In terms of phase space variables the conserved current components
 may be  written as
$$
\eqalignno{
(\0j )_{ij} &= \varphi_i \pi_j - \varphi_j\pi_i \quad ,&(13a)\cr
(\1j )_{ij} &= \varphi_i \varphi'_j - \varphi_j\varphi'_i\quad . & (13b)
\cr}
$$
Notice that $j_\mu$ is an antisymmetric matrix-valued field.
On the other hand the intertwiner field
 given in (7) is symmetric,
$$
(j)_{ij} = \varphi_i\varphi_j \quad .\eqno(13c)
$$
We observe that the Hamiltonian (10) can be written in
the Sugawara form
$$
{\cal H} = -{1\over 4}  \tr ( j_0^2 + j_1^2)  \quad .\eqno(14)
$$
It is convenient to present the current algebra in terms of matrix components,
which follows from the elementary brackets (12):
$$
\eqalign{
\{ (\0j )_{ij}(x) , (\0j )_{kl}(y) \} &= (\delta \circ  \0j )_{ij,kl}(x)
\delta(x-y)\cr
\{ (\1j )_{ij}(x) , (\0j )_{kl}(y) \} &= (\delta \circ   \1j )_{ij,kl}(x)
\delta(x-y) + (\delta \circ   j)_{ij,kl}(x) \delta'(x-y)\cr
\{ (\1j )_{ij}(x) , (\1j )_{kl}(y) \} &= 0\cr
{}&{}\cr
\{ (j)_{ij}(x) , (j)_{kl}(y) \} &= 0\cr
\{ (j)_{ij}(x) , (\1j )_{kl}(y) \} &= 0\cr
\{ (j)_{ij}(x) , (\0j )_{kl}(y) \} &= -(\delta \star j)_{ij,kl}(x)
\delta(x-y)\cr} \eqno(15)
$$
where
$$
\eqalignno{
(\delta\circ   A)_{ij,kl}&\equiv \delta_{ik}A_{jl} - \delta_{il}A_{jk} +
\delta_{jl}A_{ik} - \delta_{jk}A_{il}&(16)\cr
(\delta\star  A)_{ij,kl}&\equiv \delta_{ik}A_{jl} - \delta_{il}A_{jk} -
\delta_{jl}A_{ik} + \delta_{jk}A_{il}\quad .&(17)\cr}
$$
Further useful properties of the product defined in (16) are listed in the
Appendix. The algebra of components (8) can be easily rederived from (15)
using the property (A.8).

\vskip 1.5truecm
\penalty-300
\noindent  {\bf 3. Standard  and  Improved  Non-local Charges}
\vskip 1.truecm
\nobreak
\noindent Non-local charges may be generated by a very simple
algorithm [16], starting out of a current $j_\mu$ obeying
$$
\eqalign{
\partial ^\mu j_\mu &= 0\cr
\partial _\mu j_\nu - \partial _\nu j_\mu + 2[j_\mu ,j_\nu ]&=0 \quad .\cr}
\eqno(18)
$$
Given a conserved current $J^{(n)}_\mu$, one defines the associated non-local
potential $\chi^{(n)}$ through the equation
$$
J_\mu ^{(n)} = \epsilon _{\mu \nu}\partial ^\nu \chi^{(n)}
\quad ,\eqno(19)
$$
and build the $(n+1)^{\rm th}$ order non-local current
$$ \eqalign{
J^{(n+1)} _\mu & \equiv D_\mu \chi^{(n)} = \partial _\mu \chi ^{(n)} + 2[j_\mu
,\chi ^{(n)}]\quad ,\cr
\hat Q^{(n)} & \equiv \int d x J_0^{(n)}\quad .\cr}
\eqno(20)
$$
Such current is also
conserved as a consequence of Eqs. (18).
Here we have to mention that for the first non-local current $J_\mu^{(1)}$ the
coefficient in front of commutator in (20)
must be taken as 1 instead of 2.

Eq. (19) can be
inverted for  $ \chi^{(n)} =\partial ^{-1}J_0^{(n)} $, where we choose the
antiderivative operator as
$$
\dum A(x) ={1\over 2}\int d y \, \epsilon(x-y)A(y)
\quad ,\quad \epsilon(x)= \cases{-1,\,& $x<0$\cr \phantom{-}0,\, &$x=0$\cr
+1,\, &$x>0$\cr}
\quad .\eqno(21)
$$
With this definition we have antisymmetric boundary conditions for
$\chi^{(n)}$,
$$
\chi^{(n)} (\pm \infty) = \pm {1\over 2}\int dx J_0^{(n)}(x) =\pm
{1\over 2}\hat Q^{(n)}\quad .\eqno(22)
$$

Other boundary conditions could be used  [13] but the above
choice guarantees that the algebra of charges produces antisymmetric
combinations of charges, which  belong to the $O(N)$
algebra; moreover, the charge algebra closes in terms of $O(N)$ generator.
 We call the charges defined in Eq. (20) the {\it standard} charges. After
some partial integrations the first few of them read
$$
\displaylines{
\quad\hat Q^{(0)} = \int dx \,  \0j  \hfill(23)\cr
\quad\hat Q^{(1)} = \int dx \left(\1j  + 2\0j  \dum \0j \right)\hfill(24)\cr
\quad\hat Q^{(2)} = \hat Q^{(0)}+{1\over 2}\left(\hat Q^{(0)}\right)^3+3\int
dx
\left( \1j \dum \0j  - \dum \0j \1j - 2 \dum \0j \0j \dum \0j \right)
\hfill(25)\cr
\quad\hat Q^{(3)} = \hat Q^{(1)} + 2 \hat Q^{(1)} \left(\hat Q^{(0)}\right)^2 +
4\hat Q^{(0)} \hat Q^{(1)} \hat Q^{(0)} + 2\left(\hat Q^{(0)}\right)^2
\hat Q^{(1)} \hfill \cr
\hfill+ 4\!\! \int \!\!d x\!\!\left[ \0j \dum \0j + \1j \dum \1j -2
(\dum \0j \1j \dum \0j +
\dum \0j \0j \dum \1j + \dum \1j \0j \dum \0j ) +
4 \dum (\dum \0j \0j )\0j \dum \0j \right]
\,.\,\,(26)\cr}
$$

However, it turns out that the algebra satisfied by these standard set of
charges is not transparent enough [12-14].
In the search for a more suitable basis of
charges we find out an algebraic algorithm, where the charge $\hat Q^{(1)}$
plays a fundamental role, which generates an improved set of conserved charges
($\{ Q^{(n)}\}$). We can relate this new set to the standard one: for instance,
 the first few {\it improved} charges read
$$
\eqalign{
Q^{(0)} & \equiv \hat Q^{(0)} \cr
Q^{(1)} & \equiv \hat Q^{(1)} \cr
Q^{(2)} & \equiv {2\over 3}\hat Q^{(2)} + {4\over 3}\hat Q^{(0)} -
{1\over 3}\left(\hat Q^{(0)}\right)^3 \cr
Q^{(3)} & \equiv {1\over 3}\hat Q^{(3)} + {8\over 3}\hat Q^{(1)} -
{2\over 3} \hat Q^{(1)} \left(\hat Q^{(0)}\right)^2 - {4\over 3}\hat Q^{(0)}
\hat Q^{(1)} \hat Q^{(0)} - {2\over 3} \left(\hat Q^{(0)}\right)^2 \hat Q^{(1)}
\quad . \cr}\eqno(27)
$$

In terms of the local currents $j_\mu$, we write down the first six improved
charges,
$$
\eqalign{
 Q^{(0)} &= \int dx \,  \0j  \cr
 Q^{(1)} &= \int dx \left(\1j  + 2\0j  \dum \0j \right)\cr
 Q^{(2)} &= \int dx \left( 2\0j  + 2 \1j \dum \0j - 2 \dum \0j  \1j  -
4 \dum \0j  \0j  \dum \0j \right)\cr
Q^{(3)} &=\int d x\left[ 3\1j  + 8\0j  \dum \0j  + 2 \1j \dum \1j
- 4(\dum \0j \1j \dum \0j  + \dum \0j  \0j  \dum \1j  + \dum \1j \0j \dum \0j
)\right.\cr
&\left. + 8 \dum (\dum \0j  \0j ) \0j  \dum \0j  \right] \cr
Q^{(4)} &=  \int d x \left\{ 6\0j  + 10\1j \dum \0j  - 10 \dum \0j  \1j
- 24 \dum \0j  \0j  \dum \0j  \right.\cr
&- 4(\dum \1j \0j  \dum \1j  + \dum \0j \1j \dum \1j +\dum \1j \1j \dum \0j )
\cr
 & + 8[\dum (\dum \0j \0j )(\0j \dum \1j  + \1j \dum \0j ) -(\dum \0j  \1j  +
\dum \1j  \0j ) \dum (\0j  \dum \0j )]  \cr
&\left. +16 \dum (\dum \0j  \0j ) \0j \dum (\0j \dum \0j )\right\}\cr
Q^{(5)} &=  \int d x \left\{ 10\1j  + 32\0j \dum \0j + 12\1j \dum \1j
\right.\cr
& - 28(\dum \0j \1j \dum \0j + \dum \0j \0j \dum \1j  + \dum \1j \0j \dum \0j )
- 4\dum \1j \1j \dum \1j \cr
& +64 \dum (\dum \0j \0j ) \0j \dum \0j  + 8[\dum (\dum \1j \0j ) \0j \dum \1j
+\dum (\dum \0j \0j ) \1j \dum \1j + \dum (\dum \0j \1j ) \0j \dum \1j \cr
& +\dum (\dum \1j \0j ) \1j \dum \0j + \dum (\dum \0j \1j ) \1j \dum \0j +
\dum (\dum \1j \1j ) \0j \dum \0j ] \cr
& +16[\dum (\dum \0j \0j ) \0j \dum (\0j \dum \1j ) +
      \dum (\dum \0j \0j ) \0j \dum (\1j \dum \0j ) +
       \dum (\dum \1j \0j ) \0j \dum (\0j \dum \0j ) \cr
& \left.  + \dum (\dum \0j \1j ) \0j \dum (\0j \dum \0j ) +
     \dum (\dum \0j \0j ) \1j \dum (\0j \dum \0j )] -
 32 \dum (\dum (\dum \0j \0j ) \0j )\0j \dum (\0j \dum \0j )\right\}.\cr}
\eqno(28)
$$

Using the above definitions of the improved charges and the current algebra
given in  (15) we obtain after  a rather tedious calculation the following
algebra
$$
\eqalign{
\{ Q^{(0)}_{ij}, Q^{(0)}_{kl}\} =& \left(\delta\circ Q^{(0)}\right)_{ij,kl}\cr
\{ Q^{(1)}_{ij}, Q^{(0)}_{kl}\} =& \left(\delta\circ Q^{(1)}\right)_{ij,kl}\cr
\{ Q^{(1)}_{ij}, Q^{(1)}_{kl}\} =& \left(\delta\circ Q^{(2)}\right)_{ij,kl}
- \left({Q^{(0)}}Q^{(0)} \circ Q^{(0)}\right)_{ij,kl}\cr
\{ Q^{(2)}_{ij}, Q^{(0)}_{kl}\} =& \left(\delta\circ Q^{(2)}\right)_{ij,kl}\cr
\{ Q^{(2)}_{ij}, Q^{(1)}_{kl}\} =& \left(\delta\circ Q^{(3)}\right)_{ij,kl}
 - \left( {Q^{(0)}}Q^{(0)}\circ   Q^{(1)}\right)_{ij,kl} -
\left( Q^{(1)} Q^{(0)} \circ   Q^{(0)}\right)_{ij.kl}\cr
\{ Q^{(3)}_{ij}, Q^{(0)}_{kl}\} =& \left(\delta\circ Q^{(3)}\right)_{ij,kl}\cr
\{ Q^{(3)}_{ij}, Q^{(1)}_{kl}\} =& \left( \delta\circ Q^{(4)}\right)_{ij,kl}
- \left( {Q^{(0)}}Q^{(0)}\circ Q^{(2)}\right)_{ij,kl}
- \left( Q^{(1)} Q^{(0)} \circ   Q^{(1)}\right)_{ij.kl}  \cr
& \phantom {\left( \delta\circ Q^{(4)}\right)_{ij,kl}
- \left( {Q^{(0)}}Q^{(0)}\circ Q^{(2)}\right)_{ij,kl} }
- \left( Q^{(2)}Q^{(0)}\circ  Q^{(0)}\right)_{ij,kl}\cr
 \{ Q^{(2)}_{ij}, Q^{(2)}_{kl}\} =& \left(\delta\circ Q^{(4)}\right)_{ij,kl}
- \left({Q^{(0)}}Q^{(0)} \circ Q^{(2)}\right)_{ij,kl} -
\left( Q^{(1)}Q^{(0)}\circ  Q^{(1)}\right)_{ij,kl}\cr
& \phantom {\left(\delta\circ Q^{(4)}\right)_{ij,kl} }
-\left( Q^{(0)}Q^{(1)}\circ  Q^{(1)}\right)_{ij,kl}-
\left( {Q^{(1)}}Q^{(1)}\circ Q^{(0)}\right)_{ij,kl} \cr
\{ Q^{(3)}_{ij}, Q^{(2)}_{kl}\} = & \left(\delta\circ Q^{(5)}\right)_{ij,kl}
- \left({Q^{(0)}}Q^{(0)} \circ Q^{(3)}\right)_{ij,kl} -
\left( Q^{(1)}Q^{(0)}\circ  Q^{(2)}\right)_{ij,kl}\cr
& \phantom{\left(\delta\circ Q^{(5)}\right)_{ij,kl} }
-\left( Q^{(2)}Q^{(0)}\circ  Q^{(1)}\right)_{ij,kl}
-\left( Q^{(0)}Q^{(1)}\circ  Q^{(2)}\right)_{ij,kl}\cr
& \phantom {\left(\delta\circ Q^{(5)}\right)_{ij,kl} }
- \left( Q^{(1)}Q^{(1)}\circ  Q^{(1)}\right)_{ij,kl} -
\left( Q^{(2)}Q^{(1)}\circ  Q^{(0)}\right)_{ij,kl}\cr
\{ Q^{(4)}_{ij}, Q^{(1)}_{kl}\} = &\left(\delta\circ Q^{(5)}\right)_{ij,kl}
- \left({Q^{(0)}}Q^{(0)} \circ Q^{(3)}\right)_{ij,kl}
-\left( Q^{(1)}Q^{(0)}\circ  Q^{(2)}\right)_{ij,kl}\cr
& \phantom {\left(\delta\circ Q^{(5)}\right)_{ij,kl}}
-\left( Q^{(2)}Q^{(0)}\circ  Q^{(1)}\right)_{ij,kl}
-\left( Q^{(3)}Q^{(0)}\circ  Q^{(0)}\right)_{ij,kl}\quad .\cr}\eqno(29)
$$

Indeed we have used the algebra above to {\it define} the improved charges: we
verify that the bracket $\{ Q^{(n)},Q^{(1)}\} $ always produces a term of the
form $(\delta \circ A)$ for some $A$, which we call  linear piece; and
other essentially different terms as $(B\circ C)$, with $B$ and $C$ different
from the identity matrix ($\delta$), coming from surface contributions,
which we refer to as the
non-linear piece. Therefore we can take $A$ as a definition of the $Q^{(n+1)}$
charge,
$$(\delta \circ Q^{(n+1)})\equiv \{ Q^{(n)},Q^{(1)}\} - (\hbox{n.l.t.})
\quad ,$$
where n.l.t. means ``non-linear terms''.
While the standard charges are defined through an integro-differential
algorithm, the improved ones are generated by an algebraic procedure (where
$Q^{(1)}$ plays the role of a ``step" generator).

The fact that all brackets $\{ Q^{(n-i)},Q^{(i)}\}$, $i=0,\cdots n$, produce
the same linear term $(\delta \circ Q^{(n)})$ means that the linear part of the
algebra is of the Kac-Moody type.

These results and observations  motivate us to write down the Ansatz
$$
\{ Q^{(m)}_{ij}, Q^{(n)}_{kl}\} = \left( \delta \circ   Q^{(n+m)}\right)_
{ij,kl} - \sum _{p=0}^{m-1}\sum _{q=0}^{n-1}
\left( Q^{(p)} Q^{(q)} \circ Q^{(m+n-p-q-2)}\right)_{ij,kl}
\eqno(30)
$$
which is the first main result of this paper. In order to prove it, we consider
a set of simplified charges which obey an isomorphic algebra, and we prove the
result (30) for this new set upon deriving a chain  algebra structure in a
sense to be defined in Sect. 4. We prove the Jacobi identity for the chain
algebra and finally project it back into the non-local charge algebra.

Let us call $n$ the order of the charge $Q^{(n)}$ and $n+m$ the order of the
bracket $\{ Q^{(n)}, Q^{(m)}\} $.  Having proved up to some order $N>1$
that the algebra is composed of a linear part
which is the Kac-Moody algebra plus some non-linear piece, the
question is whether the charges can always be defined in such a way that this
structure still holds for some linear combination of the previous charges.

In order to solve this problem, we first argue that if one solves
the linear part of the algebra
in such a way that only the highest order (or genus) term
survives (i.e., one has a single
term $f^{abc}Q^{(m+n)}_c$), the same will be true
concerning the non-linear part. Indeed, relying upon the
chain algebra (which will be introduced and discussed in Sect. 4) we
notice that, in general, a non-local charge can be defined in
terms of a group theoretical
factor times some integrals of {\it chains}, constructed with both
components of the current $j_\mu $, which we can denote as in Fig. 1 where the
full dots denote the insertion of the component $\1j $ and the empty ones
$\0j $.
The ``longest" chain of a charge constitutes its highest order term.
The Dirac brackets of two chains might generate lower order terms in the
algebra of charges; however, the presence of a lower order
term in the linear part of the algebra implies a lower order term
in the non-linear part as well, with the same coefficient.

Therefore we argue with the linear algebra only, and this must be enough. We
further suppose that when $n+m< N$ for some $N>1$
the linear part is of the Kac-Moody type (as we verified in (29) for $N=6$),
namely
$$\left\{ Q^{(n)}_a,Q^{(m)}_b\right\} =-f^{abc}Q^{(n+m)}_c
+\, (\hbox{n.l.t.})\quad .
\eqno(31)$$
Let us suppose that for $n'+m'\ge N$ the algebra is non-saturated, i.e. it
contains lower order terms,
$$\left\{ Q^{(n')}_a,Q^{(m')}_b\right\} =-f^{abc}\left[ Q^{(n'+m')}_c +
a_{n',m'}Q^{(n'+m'-1)}_c+b_{n',m'}Q^{(n'+m'-2)}_c\cdots\right] +
(\hbox{n.l.t.})
\eqno(32)$$
We consider the Jacobi identity
$$
\{\{Q_a^{(n)},Q_b^{(m)}\}, Q_c^{(p)}\} +
\{\{Q_c^{(p)},Q_a^{(n)}\}, Q_b^{(m)}\} +
\{\{Q_b^{(m)},Q_c^{(p)}\}, Q_a^{(n)}\} =0 \quad .\eqno(33)
$$
Using (32) and (31) we have
$$\eqalign{
f^{abd} \{ Q_d^{(n+m)},Q_c^{(p)}\} & + f^{cad} \{ Q_d^{(n+p)} +
a_{n,p}Q_d^{(n+p-1)} +b_{n,p}Q_d^{(n+p-2)} + \cdots,Q_b^{(m)}\} \cr
& + f^{bcd} \{ Q_d^{(m+p)} + a_{m,p}Q_d^{(m+p-1)} +
b_{m,p}Q_d^{(m+p-2)}+ \cdots ,Q_a^{(n)}\} =0 \cr} \eqno(34)
$$
which implies, upon use of the relation
$$
f^{abd}f^{dce} + f^{cad}f^{dbe} + f^{bcd}f^{dae}=0
\quad ,\eqno(35)
$$
the result
$$
a_{n+m,p} = a_{n+p,m} + a_{n,p} = a_{m+p,n}+ a_{m,p}\quad . \eqno(36)
$$
If $p<n,m$, we have $a_{n,p}=a_{m,p}=0$ by the induction hypothesis, therefore
$$
a_{n+m,p}= a_{n+p,m} = a_{m+p,n}\quad .\eqno(37)
$$
Since the l.h.s. only depends on the contribution $n+m$, we conclude that
$a_{n,m}$ only depends on $n+m$. It is a simple exercise to show that the same
is true for the coefficients $b,c$ etc, therefore the linear part of
$\{ Q_a^n,Q_b^m\}$ only depends on $n+m$, and we can redefine the r.h.s.
$$
f^{abc}\left[ Q_c^{n+m}+ a(n+m)Q_c^{n+m-1} + \cdots \right] +
(\hbox{n.l.t.}) =
f^{abc}\widetilde Q_c ^{n+m} + (\hbox{n.l.t.})
\eqno(38)
$$
in such a way that it is of the Kac Moody type  after the above redefinition.
Therefore there must exist a basis of charges satisfying the algebra (30).
\vskip 1.5truecm
\penalty-300
\noindent{\bf 4. Saturated Charges}
\vskip 1.truecm
\nobreak

\noindent Consider the improved basis of charges: from the examples listed in
(28) we see that each one of them has a higher order piece, containing
the maximum
number of current components (the component $\0j $) in the integrand, depicted
below,
$$
\eqalign{
 Q^{(0)} &= \int dx \,  \0j (x) \cr
 Q^{(1)} &= \cdots +2\int dx\;\0j  \dum \0j
          = \cdots +\int dxdy\; \0j (x)\epsilon (x-y)\0j (y) \cr
 Q^{(2)} &= \cdots +4\int dx \0j \dum  (\0j  \dum \0j )\cr
         &= \cdots +\int dxdydz\; \0j (x)\epsilon (x-y)\0j
            (y)\epsilon (y-z)\0j (z)\cr
Q^{(3)} &=\cdots +8\int dx\;  \0j \dum (\0j \dum ( \0j  \dum \0j )) \cr
        &=\cdots +\int dxdydzdw\; \0j (x)\epsilon (x-y)
          \0j (y)\epsilon (y-z)\0j (z)\epsilon (z-w)\0j (w) \cr
\cdots &{\hskip 4truecm} \cdots\cr
Q^{(n)} &=\cdots +\int \prod _{i=0}^n dx_i\; \0j (x_0)\epsilon
(x_0-x_1)\0j (x_1) \cdots \epsilon (x_{n-1}-x_n)\0j (x_n)\,\,.\cr }\eqno(39)
$$
Inspired by the saturated character of the algebra (29) (i.e., the presence of
highest order terms only) and the expressions
above we propose the definition of the {\it saturated} charges
$$
\sq ^{(n)}\equiv \int \prod _{i=0}^n dx_i\; {\cal J}(x_0,\cdots ,x_n)
\eqno(40)
$$
where the non-local densities
$$
{\cal J}(x_0,\cdots ,x_n)\equiv \0j (x_0)\epsilon
(x_0-x_1)\0j (x_1) \cdots \epsilon (x_{n-1}-x_n)\0j (x_n)
\eqno(41)
$$
can be seen as {\it chains} of current components $\0j (x_i)$ connected by
non-local $\epsilon $ functions. We emphasize that the saturated charges
$\overline Q^{(n)}$ are not conserved quantities. Nevertheless we prove that
they realize the algebra (30) and  use this fact to verify that the
Ansatz satisfies the Jacobi identity. The fact that Jacobi identity holds for
the general case is discussed at the end of the section.

In a given basis $\{ t^a\} $ for the $O(N)$ algebra, the components of a
saturated charge can be built up from the integral of
a linear chain of components $\0j ^a$ times a group theoretical (trace) factor:
$$\sq _a^{(n)}=-{1\over 2}{\rm tr}(t^at^{a_0}\cdots t^{a_n})
\int \prod _{i=0}^n dx_i\; {\cal J}^{a_0\cdots a_n}(x_0,\cdots ,x_n)
\eqno(42)
$$
where
$$
{\cal J}^{a_0\cdots a_n}(x_0,\cdots ,x_n)\equiv \0j ^{a_0}(x_0)\epsilon
(x_0-x_1)\0j ^{a_1}(x_1) \cdots \epsilon (x_{n-1}-x_n)\0j ^{a_n}(x_n)
\quad .\eqno(43)
$$

The algebra of saturated charges follows from the algebra of chains.
In order to understand this relation we first consider the case of the simplest
brackets and later generalize the results. Consider the first non-trivial
saturated charge matrix
$$
\sq ^{(1)}_{ij} = \int d x d y\; (\0j )_{ik}(x) \, \epsilon
(x-y)\,
(\0j )_{kj}(y)
\quad .\eqno(44)
$$
Defining the components
$$
\sq _a^{(1)} = -{1\over 2}\tr \left( t^a \sq^{(1)} \right) \quad ,
\eqno(45)
$$
we have
$$\displaylines{\hfill
\sq _a^{(1)} = -{1\over 2}\tr(t^at^bt^c)\int d x d y\,
\J ^{bc}(x,y)\hfill(46)\cr
\J ^{ab}(x,y)  = \0j ^a(x) \epsilon (x-y) \0j ^b(y)\quad . \cr}
$$

The algebra obeyed by the chains in (46) is  easily derived upon use of (8),
$$
\eqalign{
\{ \J ^{ab}(x,y) , \J ^{cd}(z,w)\}  = &\{ \0j ^a(x) \epsilon (x-y) \0j ^b(y)\,
,\,
\0j ^c(z)\epsilon (z-w) \0j ^d(w) \} \cr
=&-f^{bce}\J ^{aed}(x,y,w)\delta (y-z) + f^{bde}\J ^{aec}(x,y,z)
\delta(y-w)\cr
&+ f^{ace}\J ^{bed}(y,x,w)\delta (x-z) - f^{ade}\J ^{bec}(y,x,z)\delta(x-w)
\cr}
\eqno(47)
$$
where
$\J ^{abc} (x,y,z) = \0j ^a(x) \epsilon (x-y) \0j ^b(y) \epsilon (y-z)
\0j ^c(z)$ is a 3-current chain.
Therefore we obtain for the algebra involving the charges (45) the expression
$$
\{ \sq ^{(1)}_a, \sq ^{(1)}_b\} = -
\tr (t^at^ct^d)f^{deg}\tr (t^bt^et^f)\int d
x d y d z\, \J ^{cgf}(x,y,z)\quad .
\eqno(48)
$$
Taking a basis $\{ t^a\} $ for the $O(N)$  algebra one verifies that the traces
in (48) merge into four traces,
$$\eqalign{
&-\tr (t^at^ct^d)f^{deg}\tr (t^bt^et^f)=\cr
&= \tr (t^gt^at^ct^ft^b) - \tr (t^gt^ct^at^ft^b) -\tr (t^at^ct^gt^ft^b)
+ \tr (t^ct^at^gt^ft^b)\quad .\cr}
\eqno(49)
$$
The third  trace leads to
$$
\tr (t^at^bt^ft^gt^c )\int d xd y d z\, \J ^{cgf}(x,y,z) =
\tr \left( t^at^b \sq ^{(2)}\right)\quad, \eqno(50)
$$
which is the Kac-Moody part of the algebra (30). As for the other three terms,
we use the expression
$$\eqalign{
\J ^{abc}(x,y,z) &=\0j ^a(x) \epsilon (x-y) \0j ^b(y) \epsilon (y-z) \0j
^c(z)\cr
&= -{1\over 2}\int d w [\epsilon (y-x) \0j ^a(x)] {\partial
\over \partial y}[\epsilon (y-w)\0j ^b(w)] [\epsilon (y-z)\0j ^c(z)]\cr }
\eqno(51)
$$
and verify that upon contraction with the remaining three terms in (49) we
get the integral of a total derivative which, due to the antisymmetric
boundary condition (22), gives the cubic term
$$
-\tr \left( t^a\sq ^{(0)} \sq ^{(0)} t^b\sq ^{(0)}\right) \quad ,\eqno(52)
$$
and we arrive at
$$
\{ \sq ^{(1)}_a, \sq ^{(1)}_b\} = \tr \left( t^at^b\sq ^{(2)}\right)
- \tr \left( t^a \sq^{(0)}\sq^{(0)}t^b\sq^{(0)}\right)
\quad .\eqno(53)
$$

Further algebra, and use of the boundary condition to show the vanishing of
terms of the type $\int d x \partial \tr (At^aA^tt^b)$, lead to the next
(Dirac) brackets
$$
\{ \sq ^{(1)}_a,\sq ^{(2)}_b\} =
\tr \left( t^at^b\sq ^{(3)}\right) - \tr \left(
t^a\sq^{(0)}\sq^{(0)}t^b\sq^{(1)}\right) -
\tr \left( t^a\sq^{(0)}\sq^{(1)} t^b\sq^{(0)}\right)\quad. \eqno(54)
$$
We are now in position to generalize the procedure for  arbitrary chains and
obtain the full algebra of the saturated charges.
\eject

The 2-chain brackets can be expanded as a sum of crossing chains:
$$
\displaylines{
\{{\cal J}^{a_0\cdots a_m}(x_0,\cdots ,x_m),{\cal J}^{b_0\cdots b_n}
(y_0,\cdots ,y_n)\} =\cr
=-\int \prod _{k=0}^mdx_k\prod _{l=0}^ndy_l\times
\sum _{i=0}^m\sum _{j=0}^n \delta (x_i-y_j)\; f^{a_ib_jc}\cr
\eqalign{
&\times\0j ^{a_0}\ss{(x_0)\epsilon(x_0 - x_1)} \cdots
\0j ^{a_{i-1}}\ss{(x_{i-1})\epsilon (x_{i-1} - x_i)}
\quad\times\quad\ss{\epsilon(x_i - x_{i+1})}
\0j ^{a_{i+1}}\ss{(x_{i+1})}
\cdots
\ss{\epsilon (x_{m-1} - x_m)}\0j ^{a_m}\ss{(x_m)}\cr
&{\phantom{
\times\0j ^{a_0}\ss{(x_0)\epsilon (x_0 - x_1)} \cdots
\0j ^{a_{i-1}}\ss{(x_{i-1})\epsilon (x_{i-1} - x_i)}}}
\times\0j ^c\ss{(x_i)}\cr
&\times\0j ^{b_0}\ss{(y_{0})\epsilon ({y_{0} - y_{1}})} \cdots
\0j ^{b_{j-1}}\ss{(y_{{j-1}})\epsilon (y_{{j-1}} - x_{i})}
\quad\times\quad
\ss{\epsilon(x_{i} - y_{{j+1}})}\0j ^{b_{j+1}}\ss{(y_{{j+1}})} \cdots
\ss{\epsilon (y_{{n-1}} - y_{n})}\0j ^{b_n}\ss{(y_{n})}}\cr
\hfill(55)\cr
}
$$
A typical crossing of chains is represented by Fig. 2, followed by the
respective $f^{a_ib_jc}$ structure constant factor.

The algebra of two  linear chains has produced another kind of non-local
density: a 1-crossing chain. Computing the algebra of this extended family of
chains, one generates multiple crossings. We therefore consider the infinite
space of $n$-crossing chains, whose Dirac brackets define an infinite
dimensional linear algebra. This is the underlying linear structure behind the
non-local charge algebra, which we compute now.

In order to project the brackets above into $\{ \sq _a^{(m)},\sq _b^{(n)}\} $
we must multiply both sides of Eq. (55) by the corresponding traces of
$t$-matrices, as indicated by (42). Those traces contracted by a factor
$f^{a_ib_jc}$ merge as follows
$$
\eqalign{
& -{1\over4}
 \tr (t^{a_{i+1}}\cdots t^{a_m}t^{a}t^{a_0} \cdots t^{a_i}) f^{a_ib_jc}
\tr (t^{b_j}\cdots t^{b_n}t^b t^{b_0}\cdots t^{b_{j-1}})\cr
&= {1\over4}\tr \Big( t^c \times (t^{a_{i+1}}\cdots t^{a_m}t^{a}t^{a_0} \cdots
t^{a_{i-1}}+(-)^m
t^{a_{i-1}}\cdots t^{a_0}t^at^{a_m}\cdots t^{a_{i+1}})\cr
&\phantom{= \tr [ t^c } \times (t^{b_{j+1}}\cdots t^{b_n}t^{b}t^{b_0}
\cdots t^{b_{j-1}}+(-)^n t^{b_{j-1}}\cdots t^{b_0}t^bt^{b_n}\cdots
t^{b_{j+1}})\Big)
\quad ,\cr}\eqno(56)
$$
which generalizes Eq. (49). Once again, each contraction of chains leads to
four contributions, as in Fig. 3.
It is easy to verify that each contribution shows up four times,
as exemplified by
Fig. 4: the factors $(-)^n$ and $(-)^m$ compensate the implied inversions of
arguments of the $\epsilon $-functions. Thus we can concentrate on the 3rd
representative of Fig. 3 and drop the 1/4 factor before the trace: that
figure represents a typical partition of the chains, and the sum of all
partitions in the bracket $\{ \overline Q^{(m)}_a, \overline Q^{(n)}_b\}$ reads
$$
\displaylines{
\{ \overline Q^{(m)}_a, \overline Q^{(n)}_b\} = \int
\prod_{k=0}^m d x_k \prod_{l=0}^n d y_l \sum _{i=0}^m \sum _{j=0}^n
\delta(x_i-y_i)\times(-)^m\times (-)^{m-i-1}\times (-)^{i-1}\cr
 \times \tr [t^a {\cal J} (x_m\cdots x_{i+1}) \epsilon (x_i- x_{i+1})
\epsilon(x_i - y_{j+1}) {\cal J} (y_{j+1}\cdots y_n) t^b {\cal J} (y_0
\cdots y_{j-1} x_i x_{i-1}\cdots x_0)]\cr}
$$
\eject
$$
\displaylines{
= \int \prod_{k=0}^m d x_k \prod_{l=0}^n d y_l
 \sum _{i=0}^m\sum _{j=0}^n \delta (x_i-y_j)\cr
\times\tr \!\!\Big[ t^a [\epsilon (x_i\!- \!x_{i+1}) {\cal J} (x_m\cdots
x_{i+1}\!)]
[\epsilon (x_i\! - \!y_{j+1}\!) {\cal J} (y_{j+1}\cdots y_n\!)] t^b [{\cal J}
(y_0\cdots y_{j-1} x_i x_{i-1}\cdots x_0\!)]\Big]\cr
= \int \tr [t^at^b {\cal J}(y_0\cdots
y_{n-1}x_mx_{m-1}\cdots x_0)]  \cr
- \sum _{i=0}^{m-1}\int d x \tr \Bigg[ t^a \Big[ \int
\epsilon (x-x_{i+1}) {\cal J}(x_m \cdots x_{i+1})\Big] t^b \Big[ \int
{\cal J}(y_0\cdots y_{n-1} x x_{i-1}\cdots x_0) \Big]\Bigg]\cr
+ \sum _{j=0}^{n-1}\int d x \tr \Bigg[ t^a \Big[ \int
\epsilon (x-y_{j+1}) {\cal J}(y_{j+1} \cdots y_{n})\Big] t^b \Big[ \int
{\cal J}(y_0\cdots y_{j-1} x x_{m-1}\cdots x_0) \Big]\Bigg]\cr
- \sum _{i=0}^{m-1}\sum _{j=0}^{n-1} \int d x \tr \Bigg[ t^a \Big[ \int
\epsilon (x-x_{i+1}) {\cal J}(x_m \cdots x_{i+1})\Big] \Big[ \int \epsilon
(x-y_{j+1}){\cal J} (y_{j+1}\cdots y_n) \Big]
t^b\cr
\hfill\times \Big[ \int
{\cal J}(y_0\cdots y_{j-1} x x_{i-1}\cdots x_0) \Big]\Bigg]\quad .\hfill(57)
\cr}
$$
Above we have omitted integration measures of the labelled variables, which we
assume to be resumed under the integral symbols.

Examining the four sums above,
we recognize the Kac-Moody piece in the first term
$$
\int \tr \, (t^a t^b {\cal J} (y_0\cdots y_{n-1}
x_m x_{m-1} \cdots  x_0)) = \tr \, (t^at^b Q^{(n+m)})\quad .\eqno(58)
$$

The cubic piece of the algebra can be obtained  from the last sum in (57),
which we rewrite as follows
$$
\displaylines{
- \!\!\sum _{i=0}^{m-1}\!\sum _{j=0}^{n-1}\!\! \int\!\! d x {\partial
\over \partial x} {1\over 2} \tr\! \Big[ t^a \big[ \!\!\int \!\!
\epsilon (x\!-\!x_{{}_{i+1}}) {\cal J}(x_m \cdots x_{{}_{i+1}})\big]
\big[\!\! \int \!\!\epsilon(x\!-\!y_{{}_{j+1}}){\cal J} (y_{{}_{j+1}}
\cdots y_n) \big] t^b\hfill\cr
\times  \big[ \int \epsilon (x-x_i) {\cal J}(y_0
\cdots y_{{}_{j-1}} x_i x_{{}_{i-1}}\cdots x_0) \big] \Big]\cr
\!+\!\! \sum _{i=0}^{m-1}\!\sum _{j=0}^{n-1}\!\! \int \!\! d x_{{}_{i+1}}
 \tr \! \Big[ t^a \big[ \!\!\int \!\!{\cal J}(x_m \cdots x_{{}_{i+1}}
y_{{}_{j+1}}\cdots y_n)
\big] t^b  \big[ \!\!\int \!\!\epsilon (x_{{}_{i+1}}\!-\!x_i) {\cal J}
(y_0\cdots y_{{}_{j-1}} x_i x_{{}_{i-1}} \cdots x_0) \big] \Big]\!\!\cr
\!-\!\! \sum _{i=0}^{m-1}\!\sum _{j=0}^{n-1}\!\! \int \!\!d y_{{}_{j+1}}
 \tr \Big[ t^a \big[ \!\!\int \!\! {\cal J}(x_m \cdots x_{{}_{i+1}}
y_{{}_{j+1}}\cdots y_n)
\big] t^b  \big[ \!\!\int \!\!\epsilon (y_{j+1}\!-\!x_i) {\cal J}(y_0\cdots
y_{{}_{j-1}} x_i x_{{}_{i-1}} \cdots x_0) \big] \Big]\!\!\cr
}
$$
\eject
$$
\displaylines{
\hfill= -  \sum _{i=0}^{m-1}\sum _{j=0}^{n-1} \tr (t^a \sq^{(m-1-i)}
\sq^{(n-1-j)} t^b \sq^{(i+j)}) + \cdots\hfill\phantom{(59)}\cr
\hfill= - \sum _{i=0}^{m-1}\sum _{j=0}^{n-1} \tr (t^a \sq^{(i)} \sq^{(j)} t^b
\sq^{(m+n-i-j-2)}) + \cdots \quad .\hfill(59)\cr }
$$

The terms represented by right dots above, together with the 2nd and 3rd sums
left over in (57), may be (formally) summarized as
$$
\eqalign{
&+\!\! \sum _{i=0}^{m-1}\!\sum _{j=0}^{n}\!\! \int \!\!d x_{{}_{i+1}}
\!\!\tr \! \Big[ t^a
\big[ \!\!\int \!\!{\cal J}(\!x_m \cdots x_{{}_{i+1}} y_{{}_{j+1}}
\cdots y_n\!)\big] t^b  \big[ \!\!\int \!\!\epsilon (\!x_{{}_{i+1}}\!-\!x_i\!)
 {\cal J}(\!y_0\cdots y_{{}_{j-1}} x_i x_{i-1}\cdots x_0\!) \big]\! \Big]\cr
&- \!\!\sum _{i=0}^{m}\!\sum _{j=0}^{n-1}\!\! \int \!\!d y_{{}_{j+1}}
\!\!\tr \! \Big[ t^a
\big[ \!\!\int \!\!{\cal J}(\!x_m \cdots x_{{}_{i+1}} y_{j+1}\cdots y_n\!)\big]
t^b  \big[ \!\!\int \!\!\epsilon (\!y_{{}_{j+1}}\!-\!x_i\!) {\cal J}(\!y_0
\cdots y_{{}_{j-1}} x_i x_{{}_{i-1}}\cdots x_0\!) \big]\! \Big]\cr}\eqno(60)
$$
As exemplified by Fig. 5, the above sum generates surface terms of the form
$$
\int d x {\partial \over \partial x} \tr (t^a A(x) t^b B(x) )\eqno(61)
$$
which vanishes due to our antiperiodic boundary conditions. When $n+m$ is
odd, we also find single contributions like
$$
\int d x  \tr (t^a A(x) t^b {\partial \over \partial x} A^t(x) ) =
{1\over 2}\int d x {\partial \over \partial x} \tr (t^a A(x) t^b A^t(x))
\eqno(62)
$$
which is zero for the same reasons. Therefore we have proved that the traces
of $t$-matrices project the chain algebra into the non-local charge algebra
$$
\{ \sq ^{(m)}_a, \sq ^{(n)}_b\} = \tr \left( t^a t^b \sq ^{(m+n)}\right)
- \sum _{i=0}^{m-1}\sum _{j=0}^{n-1}
\tr \left( t^a \sq ^{(i)} \sq ^{(j)} t^b \sq ^{(m+n-i-j-2)}\right) \eqno(63)
$$

By means of the formula $(A.8)$ one recognizes that the above algebra is
isomorphic to the Ansatz $(30)$. Although this algebra has been derived for the
$O(N)$ model, one can rewrite the traces appearing on the r.h.s. of $(63)$ in
terms of the structure constants of the group and therefore generalize that
algebra for other groups.

Concerning the Jacobi identity, we begin by stressing that the above
realization of the Ansatz was built up from the elementary current component
$j_0$: as the Dirac brackets $\{j_0^a(x),j_0^b(y)\}$, given in Eq. $(8)$, obey
the Jacobi identity by hypothesis, and since the chains are defined in Eq.
$(41)$ as products of $j_0$-components, it follows that the algebra of chains
also satisfies the
Jacobi identity. On the other hand, the saturated charges are constructed by
simple integrations and linear combinations of chains, therefore implying
that the algebra $(63)$ obeys the Jabobi identity too.

If we had considered the algebra of all chains, including those having the
component $j_1$, the corresponding integrations and trace-projections would
lead us to the algebra of improved charges. The Jacobi identity of
the algebra thus obtained would follow from the algebraic
properties of $(8)$ too.
The role of the intertwiner field is marginal due to its
character as a projector.

{}From the relation between chains and saturated charges, we also understand
how
the linear and cubic parts of the algebra $(30)$ are constrained: both are
constructed from the same chains, with the same number of current components
(implying the highest order terms) and same multiplicative coefficients, as
mentioned in Sec. 3.
\vskip 1.5truecm
\noindent {\bf 5. Algebra of Non-Local Charges in WZNW Model}
\vskip 1.0truecm
\nobreak

\noindent We first reanalyze  the current algebra for the
principal chiral model with a Wess-Zumino term. This model [19] contains
  a free coupling constant $\lambda$ and, for special values of
$\lambda$, is equivalent to  the conformally invariant WZNW model while being
the ordinary chiral model for $\lambda \to 0$. Therefore, the current algebra
derived below is a generalization of the current algebras for these two
special cases. For the WZNW model, the current algebra is known to consist of
two commuting Kac-Moody algebras, while for the ordinary chiral model, it
has been presented previously.

We begin by fixing our conventions. The target space for the chiral models to
be considered here  will be a simple Lie group $G$ (which is usually,
though not necessarily, assumed to be compact) with Lie algebra ${\bf g}$, and
we use the trace ${\rm tr}$ in some irreducible representation to define the
invariant scalar product $(.\,,.)$ on ${\bf g}$, normalized so that the
long roots have length $\sqrt{\,2}$, as well as the invariant closed three-form
$\omega$ on ${\bf g}$ giving rise to the Wess-Zumino term. Explicitly, for
$\, X,Y,Z \smin {\bf g}$, we have,
$$
 (X,Y)~~=~~- \, {\rm tr}\,(XY)~~~\quad ,\eqno(64)
$$
while
$$
 \omega(X,Y,Z)~~=~~{1\over 4\pi} \; {\rm tr}\,(X[Y,Z])~~~\quad .\eqno(65)
$$
Obviously, $(.\,,.)$ and $\omega$ extend to a biinvariant metric $(.\,,.)$
on $G$ and to a biinvariant three-form $\omega$ on $G$, respectively: the
latter can alternatively be represented in terms of the left invariant
Maurer-Cartan form $\, g^{-1} dg \,$ or right invariant Maurer-Cartan form
$\, dg \, g^{-1} \,$ on $G$, as follows:
$$
 \omega~=~{1\over 12\pi} \; {\rm tr} \, (g^{-1}dg)^3~
        =~{1\over 12\pi} \; {\rm tr} \, (dg \, g^{-1})^3~~~\quad .\eqno(66)
$$
(Due to the Maurer-Cartan structure equation, this representation implies
that $\omega$ is indeed a closed three-form on $G$, and the normalization
in Eqs. (65) and (66) is chosen so that $\omega/2\pi$
generates the third de Rham cohomology group $H^3(G,\Integer)$ of $G$
over the integers, at least when $G$ is simply connected; cf.\ Ref.\
[20]. The minus sign in Eq. (64) is introduced to ensure positive
definiteness when $G$ is compact.)

In part of what follows, we work in terms of (arbitrary) local
coordinates $u^i$ on $G$, representing the metric $(.\,,.)$ by its
components $g_{ij}$ and the three-form $\omega$ by its components
$\omega_{ijk}$. Then the total action of the so called
Wess-Zumino-Novikov-Witten (WZNW) theory is the sum
$$
 S~=~S_{CH} + n S_{WZ}\quad,\eqno(67)
$$
where the action for the ordinary chiral model, $S_{CH}$ is given by (1), and
the Wess-Zumino term is
$$
 S_{WZ}~=~{1\over 6} \int_B d^3 x~\epsilon^{\kappa\lambda\mu} \,
          \omega_{ijk}(\tilde{\varphi}) \,
          \partial_\kappa \tilde{\varphi}^i \,
          \partial_\lambda \tilde{\varphi}^j \,
          \partial_\mu \tilde{\varphi}^k
       ~=~\int_B \tilde{\varphi}^\star \omega~~~\quad .\eqno(68)
$$
Here, $\varphi$ and $\tilde{\varphi}$ are the basic field and the extended
field of the theory, respectively, i.e., $\varphi$ is a (smooth) map from a
fixed two-dimensional Lorentz manifold $\Sigma$ to $G$ and $\tilde{\varphi}$
is a (smooth) map from an appropriate three-dimensional manifold $B$ to $G$,
chosen so that $\Sigma$ is the boundary of $B$ and $\varphi$ is the restriction
of $\tilde{\varphi}$ to that boundary: $\Sigma=\partial B$,
$\varphi=\tilde{\varphi} \vert_{{}_{\Sigma}}$.
The conformally invariant WZNW model is obtained at
$\, \lambda = \sqrt{\,4\pi/|n|} \,$, while the ordinary chiral model
can be recovered in the limit $\lambda\!\rightarrow\!0$. Note that if
$\omega$ were exact, we could write $\, \omega = d\alpha \,$ to obtain
$$
 S_{WZ}~=~{1\over 2} \int_\Sigma d^2 x~\epsilon^{\mu\nu} \,
          \alpha_{ij}(\varphi) \,
          \partial_\mu \varphi^i \, \partial_\nu \varphi^j
       ~=~\int_\Sigma \varphi^\star \alpha~~~\quad .\eqno(69)
$$
But of course this is not possible globally, i.e., the $\alpha_{ij}$ appearing
in this formula are neither unique nor can they be chosen so as to become the
components of a globally well-defined two-form on $G$ with respect to the
$u^i$.
Still, calculations involving quantities that arise from local variations of
the action can be performed as if this were the case, and may lead to results
that do not depend on any artificial choices. For example, recall that in the
ordinary chiral model, the canonically conjugate momenta $\pi_i$ derived from
the action $S_{CH}$ are simply given by Eq. (3) and satisfy the canonical
commutation relations (2). Similarly, in the chiral model with a Wess-Zumino
term, written in the form (69), the canonically conjugate momenta $\hat{\pi}_i$
derived from the action $S$ are given by
$$
 \hat{\pi}_i~=~\pi_i \, + \, n \, \alpha_{ij}(\varphi) \, \varphi^{\prime\,
j}\eqno(70)
$$
and satisfy the canonical commutation relations (2) with $\pi$ substituted by
$\hat \pi$. Note, however, that in contrast to the $\pi_i$, the $\hat{\pi}_i$
do not behave naturally under local coordinate transformations on $G$, so that
the canonical commutation relations (2) between the $\varphi^i$ and the
$\hat{\pi}_j$ look non-covariant. This suggests to consider instead the
commutation relations between the fields $\varphi^i$ and the $\pi_j$,
which are covariant, but exhibit non-vanishing Poisson brackets between the
momenta $\pi_i$. Indeed, it follows from (2) that
$$ \eqalign{
 \{ \pi_i(x) , \pi_j(y) \} &=  + n (\partial_i \alpha_{jk} + \partial_j
\alpha_{ki})(\varphi(x)) \varphi^{\prime k}(x) \, \delta(x-y)\cr
 & + n \partial_k \alpha_{ij}(\varphi(x)) \varphi^{\prime k}(x) \delta(x-y)
\quad,\cr}\eqno(71)
$$
\eject
\noindent
 so in the presence of the Wess-Zumino term, the commutation relations between
$\varphi^i$ and  $\pi_j$ read
$$ \eqalignno{
 \{ \varphi^i(x) , \varphi^j(y) \} & = 0 \quad ,\quad
 \{ \varphi^i(x) , \pi_j(y) \} = \delta^i_j \, \delta(x-y)\quad ,\cr
 \{ \pi_i(x) , \pi_j(y) \} & = n \omega_{ijk}(\varphi(x))
\varphi^{\prime k}(x) \delta(x-y)\quad .&(72)\cr}
$$
They are obviously covariant (all expressions behave naturally under local
coordinate transformations on $G$), since $\omega$ is a globally well-defined
three-form on $G$.

To derive the desired current algebra, we recall next that the model under
consideration has an obvious global invariance under the product group
$\, G_L \times G_R \,$, which acts on $G$ according to
$$
 g \, \longrightarrow \; (g_L,g_R) \cdot g = g_L \, g \, g_R^{-1}\quad
.\eqno(73)
$$
This action of the Lie group $\, G_L \times G_R \,$ induces a representation of
the corresponding Lie algebra $\, {\bf g}_L \oplus {\bf g}_R \,$ by vector
fields, associating to each generator $\; X = (X_L,X_R) \;$ in $\, {\bf g}_L
 \oplus {\bf g}_R \,$ the fundamental vector field $X_G$ on $G$ given by
$$
 X_G(g)~=~X_L g - g X_R~~~~~{\rm for}~g \smin G\quad .\eqno(74)
$$
As usual, invariance of the action leads to conserved Noether currents taking
values in $\, {\bf g}_L \oplus {\bf g}_R \,$ and denoted by $j_\mu$ for the
ordinary chiral model and by ${\hat j}_\mu$ for the chiral model
with a Wess-Zumino term. Explicitly, we have, for $\; X = (X_L,X_R) \;$ in
$\, {\bf g}_L \oplus {\bf g}_R \,$,
$$
 (j_\mu,X)~=~- \, {1\over \lambda^2} \, g_{ij}(\varphi) \,
                  \partial_\mu \varphi^i \, X_G^j(\varphi)\quad ,\eqno(75)
$$
while
$$
 ({\hat j}_\mu,X)~=~- \; \Big( {1\over \lambda^2} \,
       g_{ij}(\varphi) \, \partial_\mu \varphi^i \,
       + \, n \, \alpha_{ij}(\varphi) \, \epsilon_{\mu\nu} \,
       \partial\>\!^\nu \varphi^i \Big) \, X_G^j(\varphi)\quad .\eqno(76)
$$
In addition, an important role is played by the scalar field $j$ introduced
in Ref. [17], defined by
$$
 (j, X\otimes Y)~=~{1\over \lambda^2} \, g_{ij}(\varphi) \,
                   X_G^i(\varphi) \, Y_G^j(\varphi)\quad .\eqno(77)
$$

The commutation relations of the Noether currents $j_\mu$ and
${\hat j}_\mu$ under Poisson brackets can now be computed
directly. Note again, however, that in contrast to $j_\mu$,
${\hat j}_\mu$ do not behave naturally under local coordinate
transformations on $G$, so that their Dirac brackets look non-covariant.
This suggests to replace them by appropriate covariant currents $J_\mu$ which,
as it turns out, can be written entirely in terms of the Noether currents
$j_\mu$ for the ordinary chiral model (the exact definition will be given
below): it is the commutation relations of these covariant currents $J_\mu$
that form the current algebra we wish to compute (or at least an important
part thereof). The most efficient way of arriving at the desired result is
therefore to calculate, as an intermediate step, the brackets of
the Noether currents $j_\mu$, using the brackets (63); we arrive at the results
$$ \eqalign{
\{ j^a_0(x) , j^b_0(y) \} & = - f^{abc} j^c_0(x)\delta(x-y) \cr
 &+ n \omega(\varphi(x)) \left( \varphi^\prime(x) ,
          t^a_L \varphi(x) - \varphi(x) t^a_R ,
          t^b_L \varphi(x) - \varphi(x) t^b_R \right)\delta(x-y)\cr}\eqno(78)
$$
while other relations remain unchanged. The additional factors ${1\over
\lambda^2}$ have been absorbed into the normalizations of the $j_\mu$ and $j$).

Before proceeding further, we find it convenient to pass to a more
standard notation, writing $g$ and $\tilde{g}$, rather than $\varphi$ and
$\tilde{\varphi}$, for the basic field and the extended field of the theory,
respectively, and using the explicit definitions (64) of the metric
$(.\,,.)$ on $G$ and (65) of the three-form $\omega$ on $G$. Then
$$
 S_{CH}~=~- \, {1\over 2\lambda^2} \int d^2 x~\eta^{\mu\nu} \; {\rm tr}
               \left( g^{-1} \partial_\mu g \, g^{-1} \partial_\nu g
\right)\quad ,\eqno(79)
$$
while
$$
 S_{WZ}~=~{1\over 4\pi} \int_0^1 dr \int d^2 x~\epsilon^{\mu\nu} \; {\rm tr}
          \left( \tilde{g}^{-1} \partial_r \tilde{g} \,
                 \tilde{g}^{-1} \partial_\mu \tilde{g} \,
                 \tilde{g}^{-1} \partial_\nu \tilde{g} \right)\quad .\eqno(80)
$$
(Here, the extended field $\tilde{g}$ is assumed to be constant outside a
tubular neighborhood $\, \Sigma \times [0,1] \,$ of the boundary $\Sigma$
of $B$, and $r$ is the coordinate normal to the boundary.) Next, we decompose
the currents $j_\mu$ and $J_\mu$, both of which take values in $\, {\bf g}_L
\oplus {\bf g}_R \,$, into left and right currents, all of which take values in
${\bf g}\,$: $\; j_\mu = (j_\mu^L,j_\mu^R) \, , \; J_\mu =
(J_\mu^L,J_\mu^R) \,$.
Explicitly,
$$
\eqalign{
 j_\mu^L &=  - \, {1\over \lambda^2} \, \partial_\mu g \, g^{-1}\quad ,\cr
j_\mu^R &= + \, {1\over \lambda^2} \, g^{-1} \partial_\mu g \quad ,\cr}
\eqno(81)
$$
and, by definition,
$$
\eqalign{
J_\mu^L &= \left( \eta_{\mu\nu} + \alpha\epsilon_{\mu\nu} \right) j^{L\,\nu}
   = - {1\over \lambda^2} \left( \eta_{\mu\nu} +
\alpha \epsilon_{\mu\nu} \right) \partial\,^\nu g \, g^{-1}\quad ,\cr
 J_\mu^R &= \left( \eta_{\mu\nu} - \alpha
  \epsilon_{\mu\nu} \right) j^{R\,\nu} = + {1\over \lambda^2}
  \left( \eta_{\mu\nu} - \alpha
  \epsilon_{\mu\nu} \right) g^{-1} \partial\,^\nu g \quad ,\cr }
\eqno(82)
$$
where $\alpha= {n\lambda^2\over 4\pi}$. The scalar field $j$, when viewed as
taking values in the space of endomorphisms of $\, {\bf g}_L \oplus {\bf g}_R
\,$, is given by the $(2\times 2)$-block matrix
$$
 j~=~{1\over \lambda^2} \left(
                        \matrix{  1 & - {\rm Ad}(g) \cr
                          - {\rm Ad}(g)^{-1} & 1 \cr}\right)\quad .\eqno(83)
$$
In other words, for $\; X = (X_L,X_R) \;$ in $\, {\bf g}_L \oplus {\bf g}_R \,
$,
$$
 j(X)~=~{1\over \lambda^2} \bigg(
        X_L - {\rm Ad}(g) X_R \, , \, X_R - {\rm Ad}(g)^{-1} X_L \bigg)\quad
.\eqno(84)
$$
It can be shown that the covariant currents $J_\mu$ defined by
Eqs. (82) differ from the Noether currents
${\hat j}_\mu$ for the chiral model with a Wess-Zumino term
by a total curl, and that current conservation (which for both types of
currents has the same physical content, because a total curl is automatically
conserved) is identical with the equations of motion of the theory.

Now in terms of an arbitrary basis $\{ t_a\} $ of ${\bf g}$, with structure
constants $f^{abc}$ defined by $ [t^a,t^b] = f^{abc} t^c $,
the various currents are represented by their components
$$\eqalign{
 j_{\mu}^{La} &=(j_\mu,t^{La}) = - {\rm tr} (j_\mu^L \, t^a)\quad ,\cr
 j_{\mu}^{Ra} &=(j_\mu,t^{Ra}) = - {\rm tr} (j_\mu^R \, t^a)\quad ,\cr
 J_{\mu}^{La} &=(J_\mu,t^{La}) = - {\rm tr} (J_\mu^L \, t^a)\quad ,\cr
 J_{\mu}^{Ra} &=(J_\mu,t^{Ra}) = - {\rm tr} (J_\mu^R \, t^a)\quad ,\cr}
\eqno(85)
$$
and the scalar field $j$ by its components
$$
 \eta_{ab}~=~(j, t^{La}\otimes t^{Lb})~=~(j, t^{Ra}\otimes t^{Rb})~
 =~- \, {1\over \lambda^2} \, {\rm tr} \left( t^a t^a \right)\quad ,\eqno(86)
$$
$$
j^{ab}=(j, t^{La}\otimes t^{Rb})={1\over \lambda^2}{\rm tr} \left( g^{-1} t^a
g \, t^b \right)\eqno(87)
$$
where
$$
 t^{La}~=~(t^a,0)~~~,~~~t^{Ra}=(0,t^a)\quad .\eqno(88)$$
With this notation, we see that the current Dirac brackets imply the
following brackets relations for the components of the currents
$j^a_\mu$:
$$
\eqalign{
 \{ \0j ^{La}(x) , \0j ^{Lb}(y) \} &= - f^{abc} \0j ^{Lc}(x) \delta(x-y)
     + \alpha f^{abc} \1j ^{Lc}(x) \delta(x-y)\quad ,\cr
 \{ \0j ^{La}(x) , \1j ^{Lb}(y) \} &= - f^{abc} \1j ^{Lc}(x)  \delta(x-y)
     + \eta_{ab} \, \delta^\prime(x-y)\quad ,\cr
 \{ \1j ^{La}(x) , \1j ^{Lb}(y) \} &= 0\quad ,\cr
{\hfill }& {\hfill }\cr
 \{ \0j ^{Ra}(x) , \0j ^{Rb}(y) \} &= - f^{abc} \0j ^{Rc}(x)  \delta(x-y)
     - \alpha f^{abc}  \1j ^{Rc}(x) \delta(x-y)\quad ,\cr
 \{ \0j ^{Ra}(x) , \1j ^{Rb}(y) \} &= - f^{abc} \1j ^{Rc}(x)  \delta(x-y)
     + \eta_{ab} \delta^\prime(x-y)\quad ,\cr
 \{ \1j ^{Ra}(x) , \1j ^{Rb}(y) \} &= 0\quad ,\cr
{\hfill } & {\hfill } \cr
 \{ \0j ^{La}(x) , \0j ^{Rb}(y) \} &= \alpha {j'}^{ab}(x)  \delta(x-y)
\quad ,\cr
 \{ \0j ^{La}(x) , \1j ^{Rb}(y) \} &= j^{ba}(y) \delta^\prime(x-y)\quad ,
\cr
 \{ \0j ^{Ra}(x) , \1j ^{Lb}(y) \} &= j^{ab}(y) \delta^\prime(x-y)\quad ,
\cr
 \{ j_{1,a}^L(x) , j_{1,b}^R(y) \} &=  0\quad .\cr}
\eqno(89)
$$
They must be supplemented by the commutation relations between the components
of the currents $j_\mu$ and those of the field $j$.
$$
\eqalign{
 \{ \0j ^{La}(x) , j^{bc}(y) \} &= -  f^{abd}  j^{dc}(x) \delta(x-y)\quad ,
\cr
\{ \0j ^{Ra}(x) , j^{bc}(y) \} &= -  f^{acd}  j^{bd}(x) \delta(x-y)\quad ,\cr
 \{ \1j ^{La}(x) , j^{bc}(y) \} &= 0\quad ,\cr
 \{ \1j ^{Ra}(x) , j^{bc}(y) \} &= 0\quad .\cr }
\eqno(90)
$$
Finally, the components of the field $j$ commute among themselves:
$\{ j^{ab}(x) ,j^{cd}(y) \}=0$.

Using the explicit representation of the theory in terms of group
valued fields, it is very simple to check the results using the decomposition
of the momentum in terms of a local and a non-local piece as it has been done
in Ref. [5].

We are now in position to generalize the previous results for the
WZNW model.
Classically,  the equations of motion are given by the conservation laws
$$
\eqalign{
\partial _\mu \left(j^{R\mu} -\alpha \epsilon^{\mu\nu}j^R_\nu\right)&=0
\quad,\cr
\partial _\mu \left(j^{L\mu} +\alpha \epsilon^{\mu\nu}j^L_\nu\right) &=0
\quad.\cr}\eqno(91)
$$
The currents $j_\mu^{R,L}$ satisfy the zero-curvature conditions
$$
\eqalign{
\partial _\mu j^R_\nu - \partial _\nu j^R_\mu  +\lambda^2  [j^R_\mu ,j^R_\nu ]
&=0 \quad ,\cr
\partial _\mu j^L_\nu - \partial _\nu j^L_\mu  +\lambda^2 [j^L_\mu ,j^L_\nu ]
&=0 \quad .\cr}
\eqno(92)
$$

Concerning the covariant currents $J_\mu^{R,L}$ the above equations imply
$$
\eqalign{
\partial ^\mu J^{R,L}_\mu &= 0 \quad ,\cr
\partial _\mu J^{R,L}_\nu - \partial _\nu J^{R,L}_\mu + [J^{R,L}_\mu ,
J^{R,L}_\nu] &= 0\quad
.\cr}\eqno(93)
$$
The last equation states [4] that the combination  $j^R_\mu -
\alpha \epsilon _{\mu \nu} j^{R\nu}$ has also zero curvature and the
construction (19) and (20) follows immediately, replacing $j^R_0 \to J^R_0$ and
$ j^R_1 \to J^R_1 $. Moreover a similar construction holds for $J^L_\mu$,
replacing $\alpha $ by $-\alpha$. In particular, the first non-local conserved
charges read
$$
\eqalign{
Q^{R(1)}&\!=\! \int\!\! d y_1dy_2\epsilon(y_1\!-\!y_2)
(j^R_0 \!+\! \alpha j^R_1)(t,y_1) (j^R_0 \!+\! \alpha j^R_1)(t,y_2)
+2\lambda^{-2}(1\!-\!\alpha^2)
\!\int\!\! d y j^R_1(t,y)\,,\cr
Q^{L(1)}&\!=\! \int\!\! d y_1dy_2\epsilon (y_1\!-\!y_2)
(j^L_0 \!-\! \alpha j^L_1)(t,y_1)(j^L_0 \!-\! \alpha j^L_1)(t,y_2)
+2\lambda^{-2}(1\!-\!\alpha^2)
\!\int\!\! d y j^L_1(t,y)\,.\cr}
\eqno(94)
$$

The construction of previous sections can thus be performed for the WZ case
with few modifications. The chain algebra construction is not touched as well
as the saturated charge, but with the above replacement of currents. In this
way we need to use the following Dirac bracket for the current $J_0$ (written
in matrix components)
$$
\{ (J^L_0)_{ij}(x), (J^L_0)_{kl}(y) \} = (\delta\circ J^L_0)_{ij,kl}(x)
\delta (x-y)+ \alpha (\delta \circ \delta ) _{ij,kl}\delta' (x-y)\eqno(95)
$$
and we are led to the following Dirac brackets for the first few (left sector)
charges
$$
\eqalign{
\{ Q^{(0)}_{ij}, Q^{(0)}_{kl}\} =&
\phantom{4\alpha}\left(\delta\circ Q^{(0)}\right)_{ij,kl}\cr
\{ Q^{(n)}_{ij}, Q^{(0)}_{kl}\} =&
\phantom{4\alpha}\left(\delta\circ Q^{(n)}\right)_{ij,kl} +
4\alpha \left( \delta\circ Q^{(n-1)}\right)_{ij,kl}  \cr
\{ Q^{(1)}_{ij}, Q^{(1)}_{kl}\} =&
\phantom{4\alpha}\left(\delta\circ Q^{(2)}\right)_{ij,kl}
-\phantom{4\alpha}\left({Q^{(0)}}Q^{(0)} \circ Q^{(0)}\right)_{ij,kl}
 + 4\alpha \left( \delta \circ Q^{(1)}\right)_{ij,kl}\cr
\{ Q^{(2)}_{ij}, Q^{(1)}_{kl}\} =&
\phantom{4\alpha}\left(\delta\circ Q^{(3)}\right)_{ij,kl}
-\phantom{4\alpha} \left( {Q^{(0)}}Q^{(0)} \circ Q^{(1)}\right)_{ij,kl}
-\phantom{4\alpha} \left( Q^{(1)} Q^{(0)} \circ   Q^{(0)}\right)_{ij.kl} \cr
+& 4 \alpha \left( \delta \circ Q^{(2)} \right)_{ij,kl}
 - 4 \alpha \left( {Q^{(0)}}Q^{(0)}\circ Q^{(0)} \right)_{ij,kl} \cr
\{ Q^{(3)}_{ij}, Q^{(1)}_{kl}\} =&
\phantom{4\alpha}\left( \delta\circ Q^{(4)}\right)_{ij,kl}
-\phantom{4\alpha} \left( {Q^{(0)}}Q^{(0)}\circ Q^{(2)}\right)_{ij,kl}
-\phantom{4\alpha} \left( Q^{(1)} Q^{(0)} \circ   Q^{(1)}\right)_{ij.kl}  \cr
& \phantom {\phantom{4\alpha}\left( \delta\circ Q^{(4)}\right)_{ij,kl}
- \phantom{4\alpha}\left( {Q^{(0)}}Q^{(0)}\circ Q^{(2)}\right)_{ij,kl} }
-\phantom{4\alpha} \left( Q^{(2)}Q^{(0)}\circ  Q^{(0)}\right)_{ij,kl}\cr
+& 4 \alpha \left( \delta \circ Q^{(3)} \right)_{ij,kl} -4\alpha
\left( {Q^{(0)}}Q^{(0)}\circ Q^{(1)} \right)_{ij,kl} -4\alpha
\left( Q^{(1)} Q^{(0)} \circ   Q^{(0)}\right)_{ij.kl}\cr
\{ Q^{(2)}_{ij}, Q^{(2)}_{kl}\} =&
\phantom{4\alpha}\left(\delta\circ Q^{(4)}\right)_{ij,kl}
- \phantom{4\alpha}\left({Q^{(0)}}Q^{(0)} \circ Q^{(2)}\right)_{ij,kl}
-\phantom{4\alpha} \left( Q^{(1)}Q^{(0)}\circ  Q^{(1)}\right)_{ij,kl}\cr
&\phantom{\phantom{4\alpha}\left(\delta\circ Q^{(4)}\right)_{ij,kl}}-
\phantom{4\alpha}\left( Q^{(0)}Q^{(1)}\circ  Q^{(1)}\right)_{ij,kl}-
\phantom{4\alpha}\left( {Q^{(1)}}Q^{(0)}\circ Q^{(0)}\right)_{ij,kl} \cr
+& 4\alpha \left( \delta\circ Q^{(3)}\right)_{ij,kl}
- 4\alpha \left( {Q^{(0)}}Q^{(0)} \circ Q^{(1)}\right)_{ij,kl}
- 4\alpha \left( Q^{(0)} \circ Q^{(0)}Q^{(1)}\right)_{ij,kl} \cr
\{ Q^{(3)}_{ij}, Q^{(2)}_{kl}\} = &
\phantom{4\alpha}\left(\delta\circ Q^{(5)}\right)_{ij,kl}
- \phantom{4\alpha}\left({Q^{(0)}}Q^{(0)} \circ Q^{(3)}\right)_{ij,kl} -
\phantom{4\alpha}\left( Q^{(1)}Q^{(0)}\circ  Q^{(2)}\right)_{ij,kl}\cr
& \phantom{\phantom{4\alpha} \left(\delta\circ Q^{(4)}\right)_{ij,kl}}
-\phantom{4\alpha}\left( Q^{(2)}Q^{(0)}\circ  Q^{(1)}\right)_{ij,kl}
-\phantom{4\alpha}\left( Q^{(0)}Q^{(1)}\circ  Q^{(2)}\right)_{ij,kl}\cr
& \phantom {\phantom{4\alpha}\left(\delta\circ Q^{(5)}\right)_{ij,kl} }
-\phantom{4\alpha} \left( {Q^{(1)}}Q^{(1)}\circ  Q^{(1)}\right)_{ij,kl} -
\phantom{4\alpha}\left( Q^{(2)}Q^{(1)}\circ  Q^{(0)}\right)_{ij,kl}\cr
+&4\alpha \left(\delta\circ Q^{(4)}\right)_{ij,kl} -4\alpha
\left( {Q^{(1)}}Q^{(0)}\circ  Q^{(0)}\right)_{ij,kl} - 4\alpha
\left( Q^{(1)}Q^{(0)}\circ  Q^{(1)}\right)_{ij,kl}\cr
&\phantom{4\alpha \left(\delta\circ Q^{(4)}\right)_{ij,kl}}
-4\alpha \left( Q^{(0)}Q^{(1)}\circ  Q^{(1)}\right)_{ij,kl}-4\alpha
\left( {Q^{(0)}}Q^{(0)}\circ  Q^{(2)}\right)_{ij,kl}\cr}\eqno(96)
$$
Therefore we write the following Ansatz for the algebra (we suppose $m\ge n$
with no loss of generality)
$$\eqalign{
\{ Q^{(m)}_{ij}, Q^{(n)}_{kl}\}  = \left( \delta \circ   Q^{(n+m)} \right)_
{ij,kl} -& \sum _{p=0}^{m-1}\sum _{q=0 }^{n-1}\left( Q^{(p)} Q^{(q)} \circ
Q^{(m+n-p-q-2)}\right)_{ij,kl}\cr
+ 4\alpha \Biggr( \left( \delta \circ Q^{(n+m-1)}\right)_{ij,kl}
-&\sum _{p=0}^{m-2}\sum _{q=0 }^{n-1}\left( Q^{(p)} Q^{(q)} \circ
Q^{(m+n-p-q-3)}\right)_{ij,kl}\Biggl)\cr} \eqno(97)
$$
or equivalently, denoting by $\{ \, , \, \}_{_{WZ}}$ the  bracket for the
Wess Zumino model and $\{ \, ,\, \}$ for previous  brackets of the chiral
model, we summarize the results by $(n\ge m)$
$$
\{ Q^{(m)}, Q^{(n)} \}_{_{WZ}}= \{ Q^{(m)}, Q^{(n)} \} + 4\alpha \{ Q^{(m-1)} ,
Q^{(n)}\}\quad .\eqno(98)
$$
Some remarks are in order now. First, concerning the chain algebra, it clearly
goes through to the Wess Zumino case. Therefore, the Jacobi identities are
valid here as well. Using them, we can perform the proof of the redefinition of
the charges in such a way that the Ansatz (97) is valid in the same way as we
did before, except for the fact that now $a_{n+m}= 4\alpha$ (see Eq. (32)).
With the argument that the linear term determines also the coefficient of the
cubic term, we arrive at the result (97) for the complete algebra. The algebra
for the right sector follows directly from (98) through $\alpha \to -\alpha$.
Also the mixed brackets $\{ Q^{L(m)}, Q^{R(n)}\} $ vanish since $\{
(J^L_0)_{ij}(x), (J^R_0)_{kl}(y)\}=0$.

\vskip 1.5truecm
\noindent {\bf 6. Conclusions}
\vskip 1.0truecm
\nobreak

\noindent We have computed the Dirac algebra of conserved non-local charges.
The result is characterized by the order $n$ of the non-local charges
$Q^{(n)}$,
which in fact can be defined in terms of its genus [21], as computed from
scattering theory. Therefore, classifying the genus, one verifies that in the
right hand side of the Dirac algebra of charges, only the highest possible
genus contributes with a non-vanishing coefficient.

We arrived at the results considering first the algebra of chains, in terms of
which one is able to prove the Jacobi identity. Further on, we prove, still
using the algebra of chains, that there is a simple relation between the linear
and the cubic part of the algebra. Using the Jacobi identity, which at this
time is known to hold, we prove that a redefinition of the charges is always
possible, in such a way that the algebra is quite simple. Furthermore, we
verified the results up to a very high order (see Eqs. $(29)$).

This result permits us to try to obtain constraints on the correlation
functions of the theory, similarly to the massive perturbation of the $k=1$ WZW
model [9]. Such problem evaded solution for several years, but with this
approach, one  should be able to accomplish such desired constraints, once one
knows a realization of charges in terms of integro-differential operators.
Indeed, for the asymptotic charges one finds such representations [5,10].

Further problems related to the role of monodromy matrices may also be obtained
once one knows the expansion of this matrix in terms of non-local charge, a
procedure in fact studied (although following the inverse way) in [4]. For
sigma models with a simple gauge group the quantum non-local charge algebra
must be the same as we have computed substituting Dirac brackets by ($-i$)
times commutators [15].

Finally, we remark that the WZNW theory presents an algebra which is analogous
to the above one. In fact, the WZNW theory has been treated by the Bethe Ansatz
[22] with results analogous in some sense to the purely bosonic case, and one
expects many similarities.

\vskip 1.5truecm
\noindent  {\bf Appendix}
\vskip .7truecm
\nobreak

\noindent In this Appendix we list some useful formulae concerning the special
product $A\circ B$ and  the constraints involving the currents
$j $ and $j_\mu $.

The product $A\circ B$ is defined as follows
$$
(A\circ B)_{ij,kl} = A_{ik}B_{jl} - A_{il}B_{jk} + A_{jl}B_{ik} - A_{jk}B_{il}
\eqno(A.1)
$$
and possesses the properties
$$ \eqalignno{
(A\circ B)_{ij,kl} & = (B\circ A)_{ij,kl}=(A^t\circ B^t)_{kl,ij} & (A.2)\cr
(A\circ B)_{ij,ka} C_{al} - (k\leftrightarrow l)& =
 (A \circ BC)_{ij,kl} + (AC \circ B)_{ij,kl}& (A.3)\cr
(A\circ B)_{ia,kl} C_{aj} - (i\leftrightarrow j)& =
 (A\circ C^t B)_{ij,kl} + (C^t A \circ B)_{ij,kl}& (A.4)\cr
C_{ia}(A\circ B)_{aj,kl} - (i\leftrightarrow j) & =
(CA \circ B)_{ij,kl} + (A\circ CB)_{ij,kl}& (A.5)\cr
A_{ia}(B\circ C)_{ab,kl}D_{bj} - (i\leftrightarrow j) & =
(AB \circ D^tC)_{ij,kl}  + (D^tB \circ AC)_{ij,kl} & (A.6)\cr
A_{ka}(B\circ C)_{ij,ab}D_{bl} - (k\leftrightarrow l) &=
(BA^t \circ CD )_{ij,kl} + (BD \circ CA^t)_{ij,kl}&(A.7)\cr}
$$

$$
{1\over 4}t^a_{ji}t^b_{lk}(A\circ B)_{ij,kl} = {\rm tr}(t^aA\, t^bB).
\eqno(A.8)
$$

Now we list the constraints among  the currents:
$$
\eqalignno{
(j_\mu \circ j_\nu)_{ij,kl} &= (j_\mu )_{ij} (j_\nu)_{kl} + (j_\nu )_{ij}
(j_\mu)_{kl} &(A.9)\cr
(j_\mu \circ j)_{ij,kl} &=0 &(A.10)\cr
(j \circ j)_{ij,kl} &=0 &(A.11)\cr
[j_\mu ,j]_+ & = j_\mu &(A.12)\cr
[j,j]_+ & = 2j &(A.13)\cr
[j_\mu , j] & = -\partial _\mu j & (A.14)\cr
(j_1\, j) & = {1\over 2}j_1 - {1\over 2}\partial _j \quad .& (A.15)\cr}
$$
We cite also an useful relation containing the antiderivative operator $\dum$
$$
(\dum A\circ A) = {1\over 2}\partial(\dum  A\circ\dum A)\quad .\eqno(A.16)
$$
\vfill \eject
\noindent {\bf Acknowledgements}
\vskip .5truecm

\noindent We would like to thank Prof. A. Salam for the hospitality at the
International Centre for Theoretical Physics, where part of this work has been
done.
\vskip 1truecm

\noindent{\bf References}
\vskip 1.0truecm

\refer[1./Pohlmeyer, K.: Integrable Hamiltonian systems and interactions
through quadratic constraints. Commun. Math. Phys. {\bf 46}, 207-221 (1976)]

\refer[2./L\"uscher, M., Pohlmeyer, K.: Scattering of massless lumps and
non-local charges in the two-dimensional classical non-linear $\sigma$-model.
Nucl. Phys. {\bf B137}, 46-54 (1978)]

\refer[3./Zamolodchikov, A.B.,  Zamolodchikov, Al.B.: Factorized S-Matrices in
two dimensions as the exact solutions of certain relativistic quantum field
theory models. Ann. Phys. {\bf 120}, 253-291 (1979)]

\refer[4./de Vega, H.J.: Field theories with an infinite number of conservation
laws and\break B\"acklund transformations in two dimensions.
Phys. Lett. {\bf 87B}, 233-236 (1979)]

\refer[5./Abdalla, E., Abdalla, M.C.B., Rothe, K.: Non-perturbative
methods  in two dimensional quantum field theory. Singapore: World Scientific
1991]

\refer[6./Belavin, A.A., Polyakov, A.M., Zamolodchikov, A.B.: Infinite
conformal symmetry in two-dimensional quantum field theory.
Nucl. Phys. {\bf B241}, 333-380 (1984)]

\refer[7./Knizhnik, V., Zamolodchikov, A.B.: Current Algebra and Wess-Zumino
model in two dimensions. Nucl. Phys. {\bf B247}, 83-103 (1984)]

\refer[8./Mussardo, G.: Off-critical statistical models: factorized scattering
theories and bootstrap program. Phys. Rep. {\bf 218}, 215-379 (1992)]

\refer[9./Abdalla, E., Abdalla, M.C.B., Sotkov, G., Stanishkov, M.: Off
critical current algebras. University of S\~ao Paulo preprint IFUSP/P-1027
(1993)]

\refer[10./L\"uscher, M.: Quantum non-local charges and absence of particle
production in two dimensional non-linear $\sigma$-model. Nucl. Phys.
{\bf B135}, 1-19 (1978)]

\refer[11./Dolan, L.: Kac-Moody algebra is hidden symmetry of chiral models.
Phys. Rev. Lett. {\bf 47}, 1371-1374 (1981)]

\refer[12./de Vega, H.J., Eichenherr, H., Maillet, J.M.: Classical and quantum
algebras of non-local charges in $\sigma$-models. Commun. Math. Phys. {\bf 92},
507-524 (1984)]

\refer[13./Barcelos-Neto, J., Das, A., Mararana, J.: Algebra of charges in the
supersymmetric non-linear $\sigma$-model. Z. Phys. {\bf 30C}, 401-405 (1986)]

\refer[14./Gomes, M. and Ha, Y.K.: Remarks on the algebra for higher non-local
charges. Phys. Rev. {\bf D28}, 2683-2685 (1983)]

\refer[15./Abdalla, E., Forger, M., Gomes, M.: On the origin of anomalies in
the quantum non-local charge for the generalized non-linear sigma models. Nucl.
Phys. {\bf B210}, 181-192 (1982) ]

\refer[16./Br\'ezin, E., Itzykson, C., Zinn-Justin, J., Zuber, J.B.: Remarks
about the existence of non-local charges in two dimensional models. Phys. Lett.
{\bf 82B}, 442-444 (1979)]

\refer[17./Forger, M., Laartz, J., Sch\"aper, U.: Current algebra of classical
non-linear sigma models. Commun. Math. Phys. {\bf 146}, 397-402 (1992)]

\refer[18./Abdalla, E., Forger, M.: Current algebra of WZNW models at and away
from criticality. Mod. Phys. Lett. {\bf 7A}, 2437-2447 (1992)]

\refer[19./Witten, E.: Non-abelian bosonization in two dimensions. Commun.
Math. Phys. {\bf 92}, 455-472 (1984)]

\refer[20./Felder, G., Gaw\c edzki, K., Kupiainen, A.: Spectra of
Wess-Zumino-Witten Models with arbitrary simple groups. Commun. Math. Phys.
{\bf 117}, 127-158 (1988)]

\refer[21./Buchholtz, D., Lopuszanski, J.T.: Non-local charges: a new concept
in quantum field theory. Lett. Math. Phys. {\bf 3}, 175-180 (1979)]

\refer[22./Polyakov, A.M., Wiegmann, P.B.: Goldstone fields in two dimensions
with multivalued actions. Phys. Lett. {\bf 141B}, 223-228 (1984)]
\vfill\eject
{\parindent=2.0truecm
\noindent {\bf Figure Captions}
\vskip 1.0truecm
\noindent \item{Fig. 1: }{Pictorial representation of chains:
empty and full circles
represent the components $j_0(x)$ and $j_1(x)$ respectively; solid lines
represent the sign function $\epsilon(x-y)$.}
\vskip 0.6truecm
\item{Fig. 2: }{Crossing in the algebra of chains: the figure represents the
term obtained from the Dirac brackets of the factors $j_0(x_i)$ and $j_0(x_j)$
of two  linear chains.}
\vskip 0.6truecm
\item{Fig. 3: }{Loops representing the four terms obtained
in the trace projection
of the crossing in Fig. 2: an $a$-labelled cross indicates the insertion of a
matrix $t^a$; the broken line is the trace line of matrices; the curly line
represents a delta function; the square indicates the omission of a component
$j_0$.}
\vskip 0.6truecm
\item{Fig. 4: }{Four equivalent loops originated from different crossings.}
\vskip 0.6truecm
\item{Fig. 5: }{Vanishing surface terms corresponding to the sums $(60)$ in the
case $m=2$, $n=3$. Fig. 5a represents the first sum, where the terms join
together in pairs (for instance, the first and the last loops) to produce a
total derivative as in Eq. $(61)$. In Fig. 5b, which corresponds to the second
sum of $(60)$, we have a similar pairing plus the single term (in the center of
the figure) which generates a surface term of the form $(62)$.}
\item{}}

\end